\documentclass{aa}
\usepackage[dvips]{graphicx}
\usepackage{natbib}
\begin{document}
   \title{External Compton emission from relativistic jets in Galactic
black hole candidates
and ultraluminous X-ray sources}
\titlerunning{Galactic black hole candidates and ultraluminous X-ray sources}

   \author{M. Georganopoulos, F. A. Aharonian, \and J. G. Kirk}
\authorrunning{ Georganopoulos et al.}

   \offprints{Markos Georganopoulos}
   \institute{Max-Planck-Institut f\"ur Kernphysik, Postfach 10 39 80, D 69029
Heidelberg, Germany}

   \date{Received Oct 16 2001 / Accepted}

   \abstract{Galactic binary systems that contain a black hole
candidate emit hard X-rays in their low luminosity mode. We show 
that this emission can be understood as due to the Compton scattering of 
photons from the companion star and/or the accretion disk by
relativistic electrons in a jet. 
The same electrons are also responsible for the radio emission.
Two sources --- XTE~J1118+480 and Cygnus~X-1 --- are modelled as
representatives of black holes with low and high luminosity companion
stars respectively.
We further show
that the ultraluminous compact X-ray sources observed in nearby
galaxies 
have the properties expected of stellar mass black holes with high luminosity
companions in which the jet is oriented close to our line of sight.
\keywords{X-rays: binaries  - stars: individual:  Cygnus X-1,      
      XTE J1118+480 -  radiation mechanisms: non-thermal }
  }
   \maketitle
%

\section{Introduction}

Several Galactic black hole candidates in
X-ray binary systems show evidence of  relativistic outflows,
and are collectively referred to as microquasars \cite{mirabel99}.
These objects emit in two different X-ray states \cite{grove98}.
In the high-soft state the spectral energy distribution peaks
at a few keV, above which there is a soft power law component.
In the low-hard state a hard
power law extends to at least $100$ keV \citep{grove98}. 
Both states have been modelled in detail
using the comptonisation of accretion disk photons from 
a hot corona containing
a hybrid plasma of thermal and nonthermal electrons --- for a review
see Poutanen (1998).
However, despite good agreement with the observed X-ray spectra, the
disk-corona models are unable to account for the 
tight temporal correlation between the low-hard X-ray state 
and the radio emission \cite{corbel00}. 

Radio jets have been detected in several microquasars \citep{mirabel99},
including Cygnus X-1, in which a one-sided 
relativistic jet has recently been resolved in observations made
during the low-hard X-ray state \citep{stirling01}.
There is increasingly strong evidence that outflows --- possibly
relativistic ones --- may be a
generic feature of the low-hard  state of microquasars \cite{fender01a}.
It has been suggested some time ago \citep{hjellming88} 
that the radio emission is synchrotron radiation  from conical jets.
Recently it has been proposed that the synchrotron emission
 might extend up to
X-ray energies, either in the extended jet 
structure \cite{atoyan99} or close to the base of the jet
\cite{markoff01}. 
Also, inverse Compton scattering by relativistic electrons in the jets 
 has previously been proposed as a mechanism for the production of
X-rays and gamma-rays 
\cite{band86,levinson96,atoyan99}.

In this letter we show that the low-hard state X-ray
emission of microquasars can be understood as Compton scattering
of photons from the companion star and/or the accretion disk 
(external Compton scattering or {\em ECS}) by  relativistic electrons in the 
persistent jet.
We apply this idea to two sources typical of systems with low and high mass companion stars,
and show that the ultraluminous compact
X-ray sources observed in nearby galaxies
\cite{makishima00} display the
properties of beamed microquasars with high mass companion stars.  
 
\section{ECS in the low-hard state \label{main}}

In the low-hard X-ray state the relativistic electrons in the jet are exposed to
the photon fields of the companion star and the
accretion disk formed around the black hole, and therefore emit
inverse Compton radiation predominantly by up-scattering these photons.
The relative importance
of the two seed photon sources depends primarily on their
energy densities at the base of the jet, where energetic particles
are injected.
Whereas the luminosities of the star and disk are in principle observable,
only upper limits to the size of the emitting region and its distance from
the black hole  can be set.
 Consequently, only lower limits on
the energy densities can be estimated.

Consider now relativistic plasma injected at a jet inlet of radius $R$.
The plasma flows with a bulk Lorentz factor
$\Gamma$ and velocity $\beta c$, where $c$ is the speed of light.
For simplicity, assume that, in the comoving frame, 
electrons are injected isotropically with a 
power-law distribution of index $p$ confined
between Lorentz factors $\gamma_1 $ and $\gamma_2$.
The environment is permeated by an isotropic monoenergetic photon field of
photon energy $\epsilon_0$ in units of $m_{ \rm e} c^2$  
and energy density $U$.
The plasma suffers inverse Compton losses until it has reached a
distance $l$ from the inlet at which the radiation field drops off significantly.  
The average time spent by electrons in this part of the jet --- the escape time
  $t_{\rm esc}$ --- is parameterised in units of the light crossing time
$t_{\rm esc}=k\; l/ c$, where $ k \ga 1$. The photon energy
density $U$ is taken to be constant within the emission region. 
The comoving photon density in the blob frame can be written as   
$U'\approx \Gamma^2 U\approx \Gamma^2  L_{\rm s} /4\pi c d^2$, 
where $L_{\rm s}$ is the luminosity of the dominant photon source and $d$ is 
the distance of the emitting plasma  from this source.
Assuming that Compton losses dominate, the Lorentz factor $\gamma_{\rm b}$ at which the cooling and escape times are equal
 is 
\begin{equation}
 \gamma_{\rm b} \approx {3 m_{\rm e} c \over \displaystyle 4 \sigma_{\rm T}  U' t_{\rm esc}}=
{3\pi m_{\rm e} c^3  d^2  \over \sigma_{\rm T}  k l  L_{\rm s} \Gamma^2}.
\label{gbreak}
\end{equation}
The spatially averaged electron energy distribution
in the comoving  frame  can be approximately written as
\begin{equation}
n'(\gamma')=\left\{\begin{array}{lc}  
\displaystyle {Q k  \over 4\pi} \gamma'^{-p} &\mbox{\hspace{0.2in} if $\gamma_{\rm 1}\leq \gamma' \leq \gamma_{\rm b}$ }  \\[.1in]
\displaystyle {Q k  \gamma_{\rm b}  \over 4\pi}  \gamma'^{-(p+1)} &\mbox{\hspace{0.2in} if $\gamma_{\rm b}\leq \gamma' \leq \gamma_{2}$ }
\end{array}\right. ,
\end{equation}
where $Q$ is a normalisation constant related to the relativistic electron 
power $P_{\rm inj}$ injected 
in the jet by
\begin{equation}
P_{\rm inj}={4\pi\over3} R^2 Q\Gamma^2  \beta m_{\rm e} c^3 { \gamma_2^{2-p}-\gamma_1^{2-p} \over 2-p}.
\end{equation}

 The spectral index of the 
external Compton spectrum  increases from $(p-1)/2$ to $p/2$ at 
the break energy $\epsilon_{\rm b}= 4 \epsilon_0 \gamma_{\rm b}^2{\cal D}^2 $,
where  ${\cal D}=1/[\Gamma(1-\beta\cos\theta)]$ is the  Doppler factor 
and  $\theta$ is the angle between the jet axis and the observer's line of sight.  
For a continuous source, the spectrum  of ECS
for energies $\epsilon_{\rm b}\leq\epsilon\leq\epsilon_{\rm max}$, where 
$\epsilon_{\rm b}=4\epsilon_0\gamma_{\rm b}^2{\cal D}^2$, 
$\epsilon_{\rm max}=4\epsilon_0\gamma_2^2{\cal D}^2$,  is \citep{georganopoulos01}
\begin{equation}
{dL \over  d\epsilon d\Omega}\approx{\cal D}^{3+p}                                      
{ Q  k \gamma_{\rm b} V \sigma_{\rm T} c U 2^{p} \over   \pi \epsilon_0
(2+p)(4+p)}
\left( {\epsilon \over \epsilon_0}\right)^{-p/2},
\label{eq:thomson}
\end{equation}
where $V=\pi R^2 l$ is the volume of the source.
For $p<2$ the luminosity per logarithmic energy interval at the peak energy
$\epsilon_{\rm peak} \approx  \epsilon_0 {\cal D}^2 \gamma_2^2$ of the 
spectral energy distribution 
given by $\epsilon \, dL/d\epsilon d\Omega$  scales  as ${\cal D}^5$: 
\begin{equation}
L_{\rm peak}=4\pi \left( {\epsilon dL \over  d\epsilon d\Omega}\right)_{\epsilon=\epsilon_{\rm peak}}\approx {3 \pi  m_{\rm}ec^3  2^p  Q R^2  \gamma_2^{2-p}   {\cal D}^5 \over  (2+p)(4+p)  \Gamma^2}. 
\end{equation}

\subsection{Microquasars with a low mass companion}

We focus now on the low-hard X-ray state of
 microquasars with a low mass companion.
These are typically systems with a 
star of K--M spectral type.
The  microquasar XTE~J1118+480 is such a system, consisting of 
a black hole of mass
$M\approx  6.5  M_\odot$  and a K7V-M0V star of luminosity
$L_{\rm s} \approx 2-6 \times 10^{32}$ erg s$^{-1}$ separated by  
$R_{\rm s}\approx 1.7 \times 10^{11}\,$cm 
\citep{mcclintock01a}. The luminosity
of the accretion disk during the active phase is
$L_{\rm d} \approx 8.6  \times 10^{35}$ erg s$^{-1}$, with a peak photon energy
 $\approx 24$ eV \citep{mcclintock01b}. 
The accretion disk is much brighter than the companion star.
Assuming that the 
emission site is located on the axis of the
accretion disk, perpendicular to the orbital plane of the binary system,
the photon energy density at the emission site is 
dominated by accretion disk photons.

 \begin{figure}
\resizebox{\hsize}{!}{\includegraphics[bb=80 50  604 554]{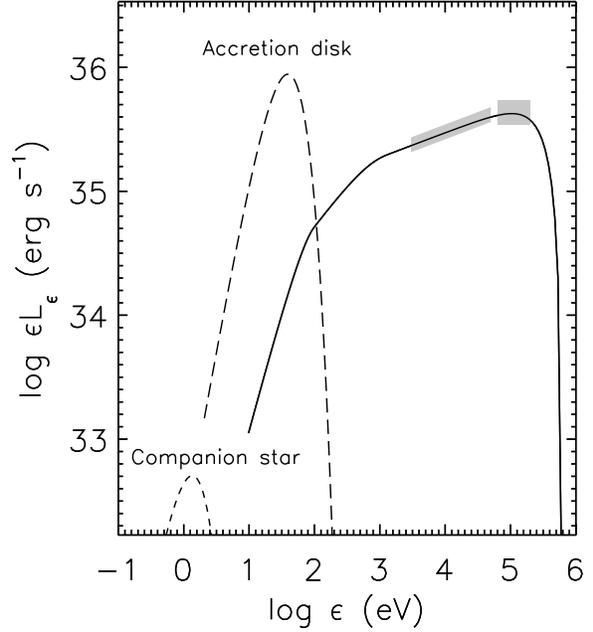}}
\caption{A fit to the X-ray emission of XTE~J1118+480, 
a system with a low mass companion. The short 
and long dashed lines represent black bodies with characteristics similar to the companion star and accretion disk
respectively. 
The solid line is the ECS emission using accretion disk seed photons. The model 
parameters are: $\Gamma=2$, ${\cal D}=1$, $P_{\rm inj}=0.05 \; L_{\rm E}$,
$d=10^8\,$cm, $R=d$, $l=2d$, $k=1$. 
The shaded area corresponds to 
observations  \protect\citep{mcclintock01b} taken in the low-hard state.}
\label{1118}
   \end{figure}

The X-ray emission has a spectral index  $\alpha \approx 0.8$ and a peak 
luminosity 
$ L_{\rm peak} \approx 4.3 \times 10^{35} $ erg s$^{-1}$ at an energy  $\approx 100$
keV. If this arises from ECS of accretion disk photons, 
the maximum energy electron is 
$\gamma_2\approx(\epsilon_{\rm peak} / \epsilon_0)^{1/2}\approx 65$ 
for a  system 
in which beaming is insignificant. 
Assuming that the observed 
spectral index $\alpha \approx 0.8$ is due to the cooled part of the electron
distribution, 
the corresponding electron injection index is $p=2\alpha=1.6$.

In this source there is  evidence for a  compact jet \citep{fender01b}
and, although its velocity in not determined, we assume, 
as  discussed in the introduction, that it is mildly relativistic,
adopting for illustration a bulk motion Lorentz factor $\Gamma=2$ and
a Doppler factor $\delta=1$, which corresponds to an angle between the
jet axis and the line of sight $\theta \approx 55 ^{\circ}$.  
To reproduce the observations we set  $d=10^8\,$cm, 
approximately  the inner radius of the truncated accretion disk 
\cite{esin01}. The jet inlet is taken to be $R=d$, and we
further set $l=2d$. 
Figure~\ref{1118} shows the spectral energy distribution, with 
injected electron power $P_{\rm inj}=0.05 L_{\rm E}$, where $L_{\rm E}$ is the 
Eddington luminosity of the black hole. The shaded area represents 
 observations  \citep{mcclintock01b}.
The solid line, corresponding to our ECS model, peaks
at $\approx 100$ keV, with a spectral index $\alpha=0.8$.

\subsection{Microquasars with a high mass companion}

\begin{figure}
\resizebox{\hsize}{!}{\includegraphics[bb=80 30 604 554]{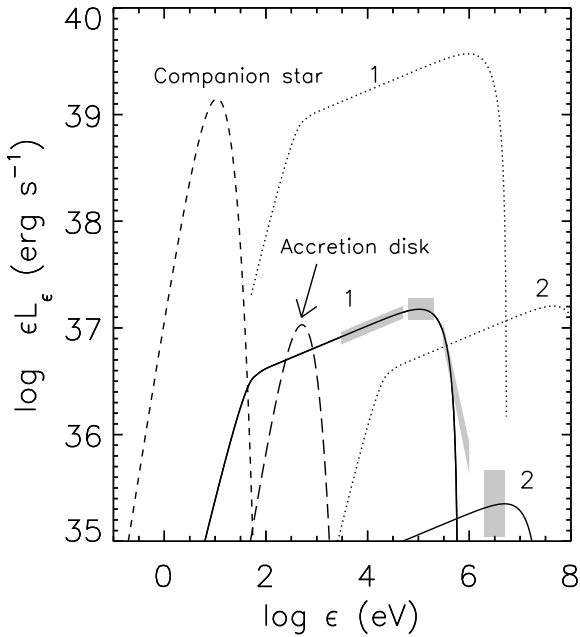}}
\caption{A fit to the X-ray emission of Cygnus X-1, a system with a
high mass companion. The short 
and long dashed lines represent  black bodies with characteristics similar to the companion star and  accretion disk
respectively. The external Compton
emission due to the star (1) and accretion disk (2) photons
is plotted for an angle $\theta$ between the jet and line of sight of
$50^{\circ}$ (solid lines)
and $10^{\circ}$ (dotted lines). The model parameters are: $\Gamma=2$, 
$P_{\rm inj}=0.5 \; L_{\rm E}$, $d_{\rm a}=3.4 \times 10^{11}\,$cm, $R=d_{\rm }a$, $l=5
R_{\rm s}$, $k=3$. 
The shaded area corresponds to observations \cite{mcconnell00b,disalvo01}.}
\label{Cygnus}
   \end{figure}

The prototype of this class is the source Cygnus~X-1. This microquasar
has a  
supergiant companion of spectral type O9.7 Iab, luminosity 
$L_{\rm s} \approx 9.6  \times 10^{38}$
erg s$^{-1}$  and a  photon energy $\approx 2.7$ eV
\citep{herrero95}.  
The separation between the companion
 star and the massive black hole (M $ \approx 10 M_\odot$) is
 $R_{\rm s} \approx 2.8 \times 10^{12}\,$cm.
Cygnus X-1 displays a soft X-ray excess in its low-hard state that 
is attributed to a truncated accretion disk. Assuming a distance of 
$2\,$kpc implies a  luminosity $L_{\rm d}\approx 7.3  \times 10^{36}\,{\rm erg\,s^{-1}}$ peaking at an 
energy $\approx 130\,$eV \citep{balucinska95}.
The X-ray spectrum peaks at $\approx 100\,$keV and has 
a luminosity 
$ L_{\rm peak} \approx 1.5 \times 10^{37}{\rm erg\,s^{-1}}$; the
spectral index  
$\alpha \approx 0.8$, corresponds to an electron injection index $p=1.6$,    \cite{mcconnell00b,disalvo01}.

The luminosity of the source at $\approx 3 $ MeV  is higher than the
 extrapolation of the spectrum from lower energies 
\citep{poutanen98, mcconnell00a}, 
possibly  suggesting a new spectral component, which nevertheless needs to be confirmed
by future spectral  measurements. 
Assuming that arises as ECS from the accretion disk photons, we obtain
the distance $d_{\rm a}$ from the base
of the jet to the accretion disk
$d_{\rm a} =3.4 \times 10^{11}\,$cm.
The distance of the jet base from the star is therefore approximately
 equal to the orbital separation, $d=R_{\rm s}$.
For modelling purposes we set $R=d_{\rm a}$
and $l=5 R_{\rm s}$. 
Under these conditions the companion star photon field dominates.
Assuming that the observed spectrum is not strongly beamed we obtain
 $\gamma_2\approx (\epsilon_{\rm peak} / \epsilon_0)^{1/2}=165$.
In fig. \ref{Cygnus} we plot the model spectral energy distribution
 for the above parameters.
The two solid lines 
correspond to the ECS emission due to the companion star (1)
and accretion disk (2) seed photons for $\theta=50^{\circ}$ ($\delta=1.12$).
Since our model has an abrupt cutoff of the electron energy distribution
at $\gamma_2$ it gives a poor fit to the very soft emission between
$100$ KeV and $1\,$MeV. 
Note, however, that  the $\sim 3\,$MeV emission 
arises naturally as the ECS of accretion seed disk photons.

\subsection{Ultraluminous X-ray sources}

Variable and, therefore, compact X-ray sources of luminosity up to 
$\sim 10^{41}\,{\rm erg\,s^{-1}}$ have been observed in nearby 
galaxies \cite{makishima00}. These sources are 
displaced from the galactic centre and, 
if they radiate isotropically, indicate a black hole mass 
$M\geq  50- 100 M_\odot$.
However, it is difficult to understand how such systems could be
formed in the required number, which has led King et al. (2001) 
to the conclusion that 
the emission may, in fact, be beamed. This immediately 
suggests a connection with the model presented above  
for Cygnus~X-1, of nonthermal ECS from a 
relativistic jet.
In the context of a synchrotron model, a similar idea has been 
taken up independently by Koerding et al. (2001). 

In the ECS model, the peak luminosity appears amplified due to
Doppler boosting by a factor of ${\cal D} ^5$ for a continuous jet. Even using the 
mildly relativistic jet parameters
appropriate for Cygnus X-1 ($\Gamma=2$, ${\cal D}_{\rm max}=4$)  
leads to a substantial maximum amplification factor of
approximately $1000$.
Such a source could reach an apparent luminosity 
$\approx 2 \times 10^{40}$ erg s$^{-1}$. 
To illustrate this point, we plot in Fig.~\ref{Cygnus} 
the spectral energy distribution predicted  
for an observer at an angle $\theta=10^{\circ}$ to the jet. 
Note that the beaming of the component due to accretion disk photons is 
weaker by a factor of 
 $(1-\beta\cos\theta)^{p/2}$
 \cite{dermer92}.

The idea that ultraluminous X-ray sources are 
related to microquasars 
such as Cygnus X-1,
is compatible with recent observations. Ultraluminous 
sources  have exhibited transitions 
from a high-soft to a 
low-hard state and vice-versa, similar to the 
spectral transitions of Galactic 
microquasars \citep{kubota01, laparola01}.
Also, optical observations suggest that the ultraluminous source NGC~5201~X-1
may have an O~star companion \citep{roberts01}
Finally, a possible X-ray periodicity has been observed in an ultraluminous
 source in  the spiral galaxy IC~342 \citep{sugiho01}, 
which, if interpreted as orbital modulation, is compatible 
with emission from a microquasar with a high mass companion star.

\section{Discussion and conclusions}

We show that the X-ray emission in the low-hard  state of microquasars
can be understood as ECS radiation from electrons in a 
relativistic jet. These electrons are 
also responsible for the radio emission occuring further downstream
 in the jet. Depending on the type of
binary system,
the ECS seed photons originate
either from the companion star and/or from the accretion disk. In our model
we assume that ECS losses dominate at the base of the jet. This
requires that the magnetic field energy density in the jet 
is lower than the external photon field energy density as seen in the frame
comoving with the flow. In the case of XTE~J118+480
this corresponds to  $B\la  10^5 $ G  whereas for Cygnus X-1
$B\la 100$ G. These upper limits
are compatible with standard  accretion disk magnetic fields,
assuming magnetic flux conservation between the accretion disk and the jet
\citep{sams96}.
In our model the 
jet is optically thick to synchrotron radiation up to $\sim$ far IR  energies;
it becomes optically thin  at $\sim$ IR  energies and the synchrotron
emission cuts off at $\sim$ UV. Thus, 
the  external  optical and UV
photons are able to penetrate the jet and act as seed photons in the
ECS mechanism.
Depending on the Thomson thickness of the outflow, a 
weak comptonisation tail due to multiple scatterings
may be  formed up to energies $\gamma_2 m c^2$. 

We have also demonstrated that the low-hard state of a Galactic
microquasar can appear as an ultraluminous X-ray source, if viewed
under a suitable angle. 
Our discussion of these sources focused on the low-hard state only.
However, by implication the soft X-ray
state must also be beamed, so that the
relativistic jet must still exists in the high-soft state.
In our model, ECS dominates the energy losses, so that it
is natural to assume that the increased accretion disk luminosity in
the high-soft state quenches the nonthermal electrons, thereby
 simultaneously producing the X-ray emission and suppressing the radio jet. 

To summarise, 
we present a model in which the X-ray emission of microquasars in the hard state
is due to external Compton scattering  by the 
energetic electrons in a relativistic jet, the same electrons being
also responsible for the  radio emission. The seed photons are stem
from the companion star and/or the accretion disk. 
We show that this model can explain the ultraluminous 
X-ray sources observed in nearby galaxies as beamed microquasars similar to 
Cygnus X-1.

\begin{acknowledgements}
      Part of this work was supported by  the European Union
TMR programme under contract FMRX-CT98-0168.
\end{acknowledgements}

\end{document}